\documentclass[epjc3]{svjour3}
\RequirePackage{graphicx}
\RequirePackage{latexsym}
\RequirePackage{amsmath}

\journalname{Eur. Phys. J. C}

\begin{document}

\title{On the effect of the degeneracy among dark energy parameters}

\author{Yungui Gong \thanksref{e1,addr1,addr2}
\and Qing Gao \thanksref{e2,addr1}}

\thankstext{e1}{e-mail: yggong@mail.hust.edu.cn}
\thankstext{e2}{e-mail: gaoqing01good@163.com}

\institute{MOE Key Laboratory of Fundamental Quantities Measurement, School of Physics, Huazhong University of Science and Technology,
Wuhan, Hubei 430074, China \label{addr1}
\and Institute of Theoretical Physics, Chinese Academy of Sciences, Beijing 100190, China \label{addr2}}

\date{Received: date / Accepted: date}

\maketitle

\begin{abstract}

The dynamics of scalar fields as dark energy is well approximated by some general relations
between the equation of state parameter $w(z)$ and the fraction energy density $\Omega_\phi$.
Based on the approximation, for slowly-rolling scalar fields,
we derived the analytical expressions of $w(z)$
which reduce to the popular
Chevallier-Polarski-Linder parametrization with explicit degeneracy relation between $w_0$ and $w_a$.
The models approximate the dynamics of scalar fields well and
help eliminate the degeneracies among $w_a$, $w_0$ and $\Omega_{m0}$. With the explicit degeneracy relations,
we test their
effects on the constraints of cosmological parameters.
We find that: (1) The analytical relations between $w_0$ and $w_a$
for the two models are consistent with observational data;
(2) The degeneracies have little effect on $\Omega_{m0}$;
(3) The $1\sigma$ error of $w_0$
was reduced about 30\% with the degeneracy relations.

\end{abstract}


\section{Introduction}

To explain the cosmic acceleration found by the observations of type Ia supernovae (SNe Ia) in 1998 \cite{hzsst98,scpsn98},
we usually introduce an exotic energy component with negative pressure to
the right hand side of Einstein equation.
This exotic energy component which contributes about 72\% to the total energy density in the universe is dubbed as dark energy.
Although the cosmological constant is the
simplest candidate for dark energy and is consistent with current
observations, other possibilities are also explored due to the many
orders of magnitude discrepancy between the theoretical estimation
and astronomical observations for the cosmological constant.
Currently we still have no idea about the nature of dark energy.  In particular,
the question whether dark energy is just the cosmological constant remains unanswered.
For reviews of dark energy, please see Refs. \cite{sahni00,peeblesde,padmanabhande,copelandde,Caldwell:2009ix,Bartelmann:2009te,limiaode}.

One way of studying the nature of dark energy is through the observational data.
There are many model-independent studies on the nature of dark energy by using the
observational data
\cite{alam04a,alam04b,barger,clarkson,corasaniti,astier,efstathiou,gerke,cooray,flux1,weller01,%
huang,star,cai,lampeitl,corray9,gong10a,gong10b,gongcqg10,gong11,Gong:2011rs,gongmnras13,li11,yu11,cai11,wetterich04,cpl1,cpl2,jbp05,zhu}.
In particular, one usually parameterizes the energy
density $\Omega_\phi(z)$ or the equation of state parameter $w(z)$ of dark energy.
Motivated by the tracking solution \cite{track1,track2} for a wide class of quintessence \cite{quintessence}
potentials in which the equation of state parameter $w(z)$
varies slowly, Efstathiou approximated $w(z)$ with $w(z)=w_0-\alpha\ln(1+z)$ in the redshift range $z< 4$ \cite{efstathiou}.
However, the most used parametrization for approximating the dynamics of a wide class of scalar fields
is the Chevallier-Polarski-Linder (CPL) parametrization
with $w(a)=w_0+w_a(1-a)$ \cite{cpl1,cpl2}.
Because of the degeneracies among the parameters $\Omega_{m0}$, $w_0$ and $w_a$ in the model,
complementary cosmological observations are needed to break the
degeneracies. The measurement on the cosmic microwave background
anisotropy, the baryon acoustic oscillation (BAO) measurement and the SNe Ia observations provide
complementary data.

On the other hand, a minimally coupled scalar field $\phi$ was often invoked to model the quintessence \cite{wetterich88,peebles88,quintessence},
and the phantom \cite{phantom}. For a scalar
field with a nearly flat potential, there exist approximate relations between the equation of state parameter $w=p/\rho$ and
the energy density parameter $\Omega_\phi$ \cite{track1,track2,Robert2008,Robert,Sourish,Crittenden:2007yy,Dutta:2008qn,Gupta:2009kk,Chiba:2009nh}.
As discussed above, the dynamics of scalar fields can be approximated with the CPL parametrization and the generic
$w-\Omega_\phi$ relations, so we expect that the degeneracies among the parameters $w_0$,
$w_a$ and $\Omega_{\phi 0}$ can be broken.
By using the generic $w-\Omega_\phi$ relations,
we can break the degeneracy between $w(z)$ and $\Omega_\phi(z)$. Furthermore, $w(z)$ can be approximated by the CPL
parametrization with $w_a$ expressed as a function of $w_0$ and $\Omega_{\phi 0}$, so the two-parameter parametrization reduces
to one-parameter parametrization \cite{gong1212}. The CPL parametrization with analytical relations among the model parameters helps
tighten the constraints on the model parameters.

In this paper, we derive two particular CPL models with $w_a$ proportional to $1+w_0$,
and study the effects of the degeneracy relations between $w_a$ and $w_0$ by using the following data:
the three year Supernova Legacy Survey (SNLS3) sample of 472 SNe Ia data with systematic errors \cite{snls3};
the BAO measurements from the 6dFGS \cite{6dfgs}, the distribution of galaxies in the Sloan Digital Sky Survey (SDSS) \cite{wjp} and the WiggleZ dark
energy survey \cite{wigglez}; the seven-year Wilkinson Microwave
Anisotropy Probe (WMAP7) data \cite{wmap7};
and the Hubble parameter $H(z)$ data \cite{hz2,hz1}.

\section{CPL parametrization with degenerated $w_0$ and $w_a$}

For a quintessence field, the equation of state parameter $w(z)$ is related with its
fractional energy density $\Omega_\phi=8\pi G\rho_\phi/3H^2$ as follows \cite{Crittenden:2007yy}
\begin{gather}
\label{womgrel1}
1+w(z)=\kappa^2(\phi)\Omega_\phi(z)(1-w(z))^2/6,\\
\label{womgrel2}
\frac{d\phi}{d\ln a}=-\kappa(\phi)\Omega_\phi(z)(1-w(z)),
\end{gather}
where $\kappa(\phi)=-3H\dot\phi/V(\phi)$.
For scalar fields satisfying the slow-roll conditions,
\begin{equation}
\label{slow1}
(\frac{1}{V}\frac{dV}{d\phi})^2\ll 1,\quad \frac{1}{V}\frac{d^2V}{d{\phi}^2}\ll 1,
\end{equation}
the dark energy density $\rho_\phi(a)$ is nearly constant and it deviates from
the cosmological constant by the order $\int_a^1(1+w)da/a$. Since $1+w(z)\ll 1$, to the zeroth order approximation,
the fractional energy density $\Omega_\phi(a)$ can be replaced by the cosmological constant
\begin{equation}
\label{phi}
\Omega_\Lambda(a)=\left[1+(\Omega_{\phi 0}^{-1}-1)a^{-3}\right]^{-1},
\end{equation}
and the dynamics of the potential can be approximated by the linear expansion of $\kappa(\phi)$
as $\kappa(\phi)=\kappa_0+\kappa_1(\phi-\phi_0)$ \cite{Crittenden:2007yy}. With these approximations,
$w(a)$ was derived as \cite{Crittenden:2007yy}
\begin{equation}
\label{waapprox22}
w(a)=-1+\frac{2}{3}\kappa_0^2\Omega_\Lambda(a)\left[\frac{\Omega_\Lambda(a)}{a^3\Omega_{\phi 0}}\right]^{2\kappa_1/3}.
\end{equation}

\begin{figure}[htp]
\centerline{\includegraphics[width=0.6\textwidth]{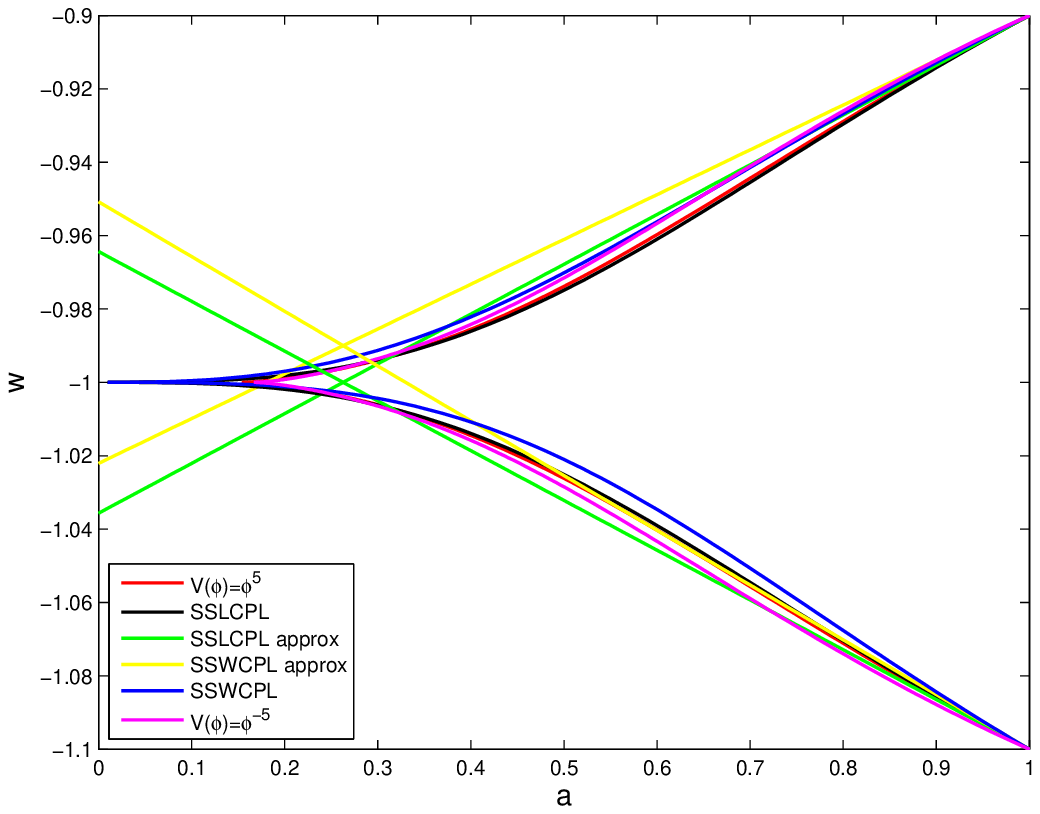}}
\caption{The evolutions of $w(a)$ for
the potentials $V(\phi)\sim \phi^5$ and $V(\phi)\sim \phi^{-5}$, the SSLCPL model (\ref{wzeq1}) and (\ref{tw})
and SSWCPL model (\ref{wzeq2}) and (\ref{tww}). The initial conditions are taken as $\Omega_{\phi0}=0.72$ and $w_0=-0.9$ for
the quintessence and $w_0=-1.1$ for the phantom. \label{waapproxfig}}
\end{figure}

On the other hand, $w(z)=\gamma-1$ also satisfies the relation \cite{Robert2008}
\begin{equation}
\label{womgrel3}
\frac{d\gamma}{d\Omega_\phi}=\frac{-3\gamma(2-\gamma)+\lambda(2-\gamma)\sqrt{3\gamma\Omega_\phi}}{3(1-\gamma)\Omega_\phi(1-\Omega_\phi)},
\end{equation}
where $\lambda(\phi)=-V^{-1}(\phi)dV(\phi)/d\phi$. For slow-roll scalar fields, $\gamma\ll 1$ and $\lambda$ is almost constant. Assume that
$\lambda(\phi)=\lambda(\phi)_{\phi=\phi_0}=\lambda_0$,
a general relationship between $w$ and the energy
density $\Omega_\phi$ for both quintessence and phantom models
was found \cite{Robert2008,Robert,Ali:2009mr},
\begin{eqnarray}
\label{eq15}
1+w=(1+w_0)\left[\frac{1}{\sqrt{\Omega_\phi}}-\left(\frac{1}{\Omega_\phi}-1\right)\tanh^{-1}(\sqrt{\Omega_\phi})\right]^2\nonumber\\
\times\left[\frac{1}{\sqrt{\Omega_{\phi0}}}-(\Omega_{\phi0}^{-1}-1)\tanh^{-1}
\sqrt{\Omega_{\phi0}}\right]^{-2}.
\end{eqnarray}
Note that the above result holds for thawing models \cite{Caldwell:2005tm} with the potentials satisfying the slow roll conditions (\ref{slow1}).
It does depend on the specific form of the potential
$V(\phi)$, furthermore, it also approximates the dynamics of tachyon fields \cite{Ali:2009mr,Chen:2013vba}.
When $w$ is close to $-1$, to the zeroth order,
the fractional energy density $\Omega_\phi$ can be approximated by the cosmological
constant $\Omega_\Lambda$ \cite{Crittenden:2007yy, Robert2008, Robert}, so
\begin{eqnarray}
\label{wzeq1}
w(a)=-1+(1+w_0)\left[\frac{1}{\sqrt{\Omega_{\phi0}}}-(\Omega_{\phi0}^{-1}-1)\tanh^{-1}
\sqrt{\Omega_{\phi0}}\right]^{-2}\nonumber\\
\times\left[\sqrt{1+(\Omega_{\phi 0}^{-1}-1)a^{-3}}-(\Omega_{\phi 0}^{-1}-1)a^{-3}\tanh^{-1}[1+(\Omega_{\phi 0}^{-1}-1)a^{-3}]^{-1/2}\right]^2.
\end{eqnarray}
In Ref. \cite{Robert2008}, it was explicitly shown in the Fig. 4 that the analytical result (\ref{wzeq1}) fits
$w(a)$ well for thawing quintessence
models with the potentials $V(\phi)\sim\phi^2$, $V(\phi)\sim\phi^{-2}$, and $V(\phi)\sim\exp(-\lambda\phi)$.
In Ref. \cite{Robert}, it was explicitly shown in the Fig. 1 that the analytical result (\ref{wzeq1})
gives the behavior of $w(a)$ for thawing phantom
models with the potentials $V(\phi)\sim\phi^6$, $V(\phi)\sim\phi^2$, $V(\phi)\sim\phi^{-2}$, and $V(\phi)\sim\exp(-\lambda\phi)$.
Therefore,  we can use $w(a)$ given by equation (\ref{wzeq1}) to approximate thawing scalar fields.
If we Taylor expand $\Omega_\phi(a)$ and $w(a)$
around $a=1$, we get
\begin{eqnarray}
\label{tphi}
\Omega_\phi\approx\Omega_{\phi 0}[1-3(1-\Omega_{\phi 0})(1-a)],
\end{eqnarray}
and
\begin{eqnarray}
\label{tw}
w=w_0+6(1+w_0)\frac{\Omega_{\phi0}^{-1/2}-\sqrt{\Omega_{\phi 0}}-(\Omega_{\phi0}^{-1}-1)\tanh^{-1}(\sqrt{\Omega_{\phi0}})}
{\Omega_{\phi 0}^{-1/2}-(\Omega_{\phi0}^{-1}-1)\tanh^{-1}(\sqrt{\Omega_{\phi0}})}(1-a).
\end{eqnarray}
Therefore, we derive the CPL parametrization with $w_a$ determined by $w_0$ and $\Omega_{\phi 0}$ starting from equation (\ref{eq15}).
We call this model as SSLCPL model.
In particular, we get \cite{gong1212,Chen:2013vba}
\begin{equation}
\label{waeq1}
w_a=6(1+w_0)\frac{(\Omega_{\phi 0}^{-1}-1)[\sqrt{\Omega_{\phi0}}-\tanh^{-1}(\sqrt{\Omega_{\phi0}})]}
{\Omega_{\phi 0}^{-1/2}-(\Omega_{\phi 0}^{-1}-1)\tanh^{-1}(\sqrt{\Omega_{\phi0}})}.
\end{equation}
When $\Omega_{\phi0}=0.7$, we get $w_a=-1.42(1+w_0)$ which is consistent with the numerical result $w\approx w_0-1.5(1+w_0)(1-a)$
obtained in \cite{Robert2008,Robert}.
Note that we derived the analytical expression of $w_a$ within the CPL parametrization which captures
the main dynamics of thawing scalar fields, this expression is not just a phenomenological dark energy
parametrization, it actually approximates the dynamics of thawing scalar fields. For the SSLCPL model,
we only have two model parameters $\Omega_{m0}$ and $w_0$ for the spatially flat case. To see how well
the approximation performs, in Fig. \ref{waapproxfig}, we show the evolutions of $w(a)$ for
the power-law potentials $V(\phi)\sim \phi^5$ and $V(\phi)\sim \phi^{-5}$, and the approximations (\ref{wzeq1}) and (\ref{tw}).
It is clear that the relative error brought by the approximation is under a few percent.
For the power-law potential $V(\phi)\sim \phi^n$ with other numbers of power $n$, the relative error is also a few percent.

From the evolution equation satisfied by $\Omega_\phi$,
\begin{equation}
\label{dynaeq2a}
\Omega'_\phi=-3w\Omega_\phi(1-\Omega_\phi),
\end{equation}
we take the approximate solution for $\Omega_\phi$ with constant equation of state $w=w_0$,
\begin{equation}
\label{omegaa}
\Omega_w=\frac{\Omega_{\phi0}a^{-3w}}{1-\Omega_{\phi0}+\Omega_{\phi0}a^{-3w}},
\end{equation}
so
\begin{eqnarray}
\label{wzeq2}
w(a)=-1+(1+w_0)\left[\frac{1}{\sqrt{\Omega_{\phi0}}}-(\Omega_{\phi0}^{-1}-1)\tanh^{-1}
\sqrt{\Omega_{\phi0}}\right]^{-2}\nonumber\\
\times\left[\sqrt{1+(\Omega_{\phi 0}^{-1}-1)a^{3w}}-(\Omega_{\phi 0}^{-1}-1)a^{3w}\tanh^{-1}[1+(\Omega_{\phi 0}^{-1}-1)a^{3w}]^{-1/2}\right]^2.
\end{eqnarray}
Then Taylor expansion $\Omega_\phi(a)$ around $a=1$ gives
\begin{equation}
\label{tphiw}
\Omega_\phi=\{1+(\Omega_{\phi 0}^{-1}-1)[1-3w(1-a)]\}^{-1}.
\end{equation}
Substituting Eq. (\ref{tphiw}) into Eq. (\ref{eq15}), we obtain,
\begin{equation}
\label{tww}
w(a)=w_0-6w_0(1+w_0)\frac{(\Omega_{\phi 0}^{-1}-1)[\sqrt{\Omega_{\phi0}}-\tanh^{-1}(\sqrt{\Omega_{\phi0}})](1-a)}
{\Omega_{\phi 0}^{-1/2}-(\Omega_{\phi 0}^{-1}-1)\tanh^{-1}(\sqrt{\Omega_{\phi0}})},
\end{equation}
so again we get CPL parametrization with
\begin{equation}
\label{twa}
w_a=-6w_0(1+w_0)\frac{(\Omega_{\phi 0}^{-1}-1)[\sqrt{\Omega_{\phi0}}-\tanh^{-1}(\sqrt{\Omega_{\phi0}})]}
{\Omega_{\phi 0}^{-1/2}-(\Omega_{\phi 0}^{-1}-1)\tanh^{-1}(\sqrt{\Omega_{\phi0}})}.
\end{equation}
We call this model as SSWCPL model.
For the SSWCPL model, we only have two model parameters $\Omega_{m0}$ and $w_0$ for the spatially flat case.
In Fig. \ref{waapproxfig}, we show the evolutions of $w(a)$ for
the approximation (\ref{wzeq2}) and (\ref{tww}).
It is clear that the relative errors brought by the approximations are under a few percent.
Contrary to the intuition that the approximation with $\Omega_w$ may be inappropriate, the numerical
results show that the relative error brought by the approximation is still small, so it is
a good approximation.
For both the SSLCPL and SSWCPL models, we find that $w_a \propto 1+w_0$, so the models are automatically consistent with $\Lambda$CDM model with $w_0=-1$ and $w_a=0$. We would like to emphasize that the models we proposed  well approximate the dynamics
of a wide class of thawing scalar fields in the whole redshift region as shown in Fig. \ref{waapproxfig},
they are different from both $\Lambda$CDM and $w$CDM model
which cannot approximate dynamical scalar fields, and they  eliminate the degeneracy between $w_0$ and $w_a$
for the CPL parametrization. Although the CPL parametrization remains to be a good approximation for
the dynamics of a wide class of scalar fields at low redshifts, the degeneracy among the model parameters is still a problem for the
fitting of cosmological data, our models break the degeneracy and help tighten the constraints on cosmological models.

\section{Observational constraints}

We apply the SNe Ia, BAO, WMAP7 and the Hubble parameter $H(z)$ data to test the effects of the degeneracy relations (\ref{waeq1}) and (\ref{twa})
on the constraints of $\Omega_{m0}$ and $w_0$.
The SNLS3 SNe Ia data consists of 123 low-redshift SNe Ia data with $z< 0.1$
mainly from Calan/Tololo, CfAI, CfAII, CfAIII and CSP,
242 SNe Ia data over the redshift range $0.08<z<1.06$ observed from the SNLS \cite{snls3},
93 intermediate-redshift SNe Ia data with $0.06< z< 0.4$ observed during the first season of SDSS-II supernova survey \cite{sdss2}, and 14 high-redshift SNe Ia data with $z > 0.8$
from Hubble Space Telescope \cite{riessgold}.
For the fitting to the SNLS3 data, we need to add two more nuisance parameters $\alpha$ and $\beta$.

The BAO data \cite{wigglez} consists of the measurement at the redshift $z=0.106$ from the 6dFGS \cite{6dfgs}, the measurements of the distribution of galaxies
at two redshifts $z=0.2$ and $z=0.35$ \cite{wjp} in the SDSS
and the measurements of the acoustic parameter at three redshifts $z=0.44$, $z=0.6$ and $z=0.73$
from WiggleZ dark energy survey \cite{wigglez}. For the BAO data,
we need to add two more nuisance parameters $\Omega_b h^2$ and $\Omega_m h^2$.

For the WMAP7 data, we use the measurements of the shift parameter
and the acoustic index at the recombination redshift \cite{wmap7}, and we
need to add two more nuisance parameters $\Omega_b h^2$ and $\Omega_m h^2$.

The Hubble parameter $H(z)$ data consists of the measurements of $H(z)$ at 11 different redshifts obtained from the differential ages of
passively evolving galaxies  \cite{hz1a,hz1}, and three data points
at redshifts $z=0.24$, $z=0.34$ and $z=0.43$, determined by taking the BAO scale as a standard ruler in the radial direction \cite{hz2}.
The $H(z)$ data spans out to the redshift regions $z=1.75$.

After obtaining the constraints on the model parameters, we reconstruct $w(z)$ and
apply the $Om$ diagnostic \cite{omztest} to detect the deviation from
the $\Lambda$CDM model. $Om(z)$ is defined as
\begin{equation}
\label{omzeq}
Om(z)=\frac{E^2(z)-1}{(1+z)^3-1},
\end{equation}
where the dimensionless Hubble parameter $E(z)=H(z)/H(z=0)$.

\begin{figure}[htp]
\centerline{\includegraphics[width=0.6\textwidth]{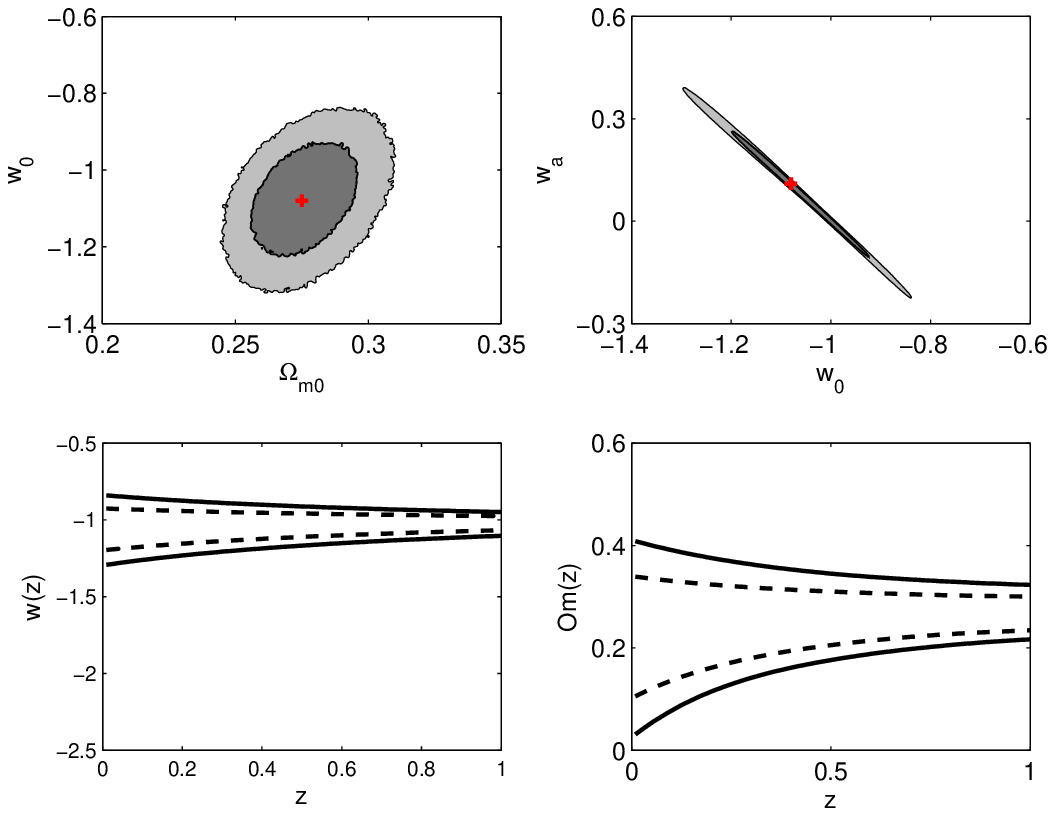}}
\caption{The marginalized $1\sigma$ and $2\sigma$ constraints on the flat SSLCPL model. The upper panels
are for the $1\sigma$ and $2\sigma$ contour plots of $\Omega_{m0}-w_0$ and $w_0-w_a$. The lower panels show the reconstructions
of $w(z)$ and $Om(z)$ by using the constraints on $\Omega_{m0}$ and $w_0$. \label{fsslcpl}}
\end{figure}

Now we consider the effects of the degeneracy relations (\ref{waeq1}) and (\ref{twa}) on $\Omega_{m0}$ and $w_0$
for the spatially flat case $\Omega_{k0}=0$. Fitting the SSLCPL model to the observational data, we
get the marginalized $1\sigma$ constraints $\Omega_{m0}=0.275^{+0.015}_{-0.011}$ and $w_0=-1.08^{+0.11}_{-0.09}$ with $\chi^2=432.6$.
By using the degeneracy relation (\ref{waeq1}) and the correlation between $\Omega_{m0}$ and $w_0$, we derived the
marginalized $1\sigma$ constraint $w_a=0.11^{+0.12}_{-0.14}$.
We show the marginalized $1\sigma$ and $2\sigma$ contours of $\Omega_{m0}$ and $w_0$, and $w_0$ and $w_a$ in Fig. \ref{fsslcpl}.
By using the correlation between  $\Omega_{m0}$ and $w_0$, we reconstruct the evolutions of $w(z)$ and $Om(z)$ in Fig. \ref{fsslcpl}.

\begin{figure}[htp]
\centerline{\includegraphics[width=0.6\textwidth]{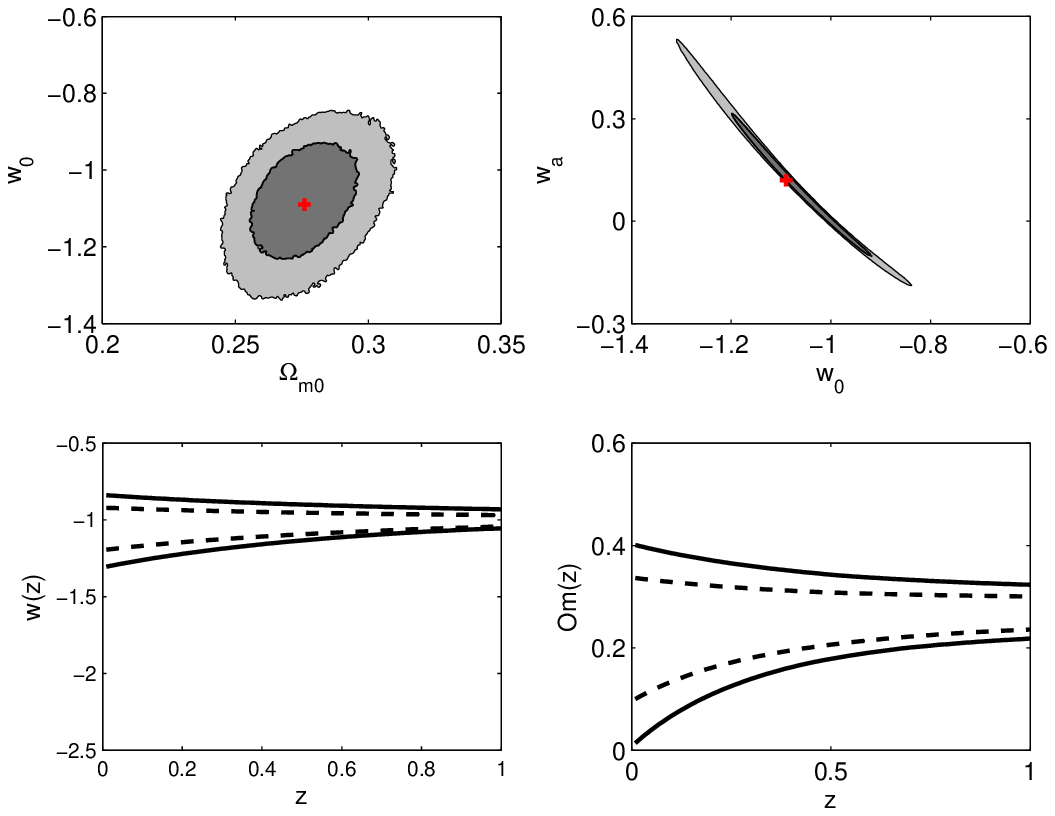}}
\caption{The marginalized $1\sigma$ and $2\sigma$ constraints on the flat SSWCPL model. The upper panels
are for the $1\sigma$ and $2\sigma$ contour plots of $\Omega_{m0}-w_0$ and $w_0-w_a$. The lower panels show the reconstructions
of $w(z)$ and $Om(z)$ by using the constraints on $\Omega_{m0}$ and $w_0$. \label{fswlcpl}}
\end{figure}

Fitting the SSWCPL model to the observational data, we
get the marginalized $1\sigma$ constraints $\Omega_{m0}=0.276^{+0.014}_{-0.013}$ and $w_0=-1.09\pm 0.10$ with $\chi^2=432.6$.
By using the degeneracy relation (\ref{twa}) and the correlation between $\Omega_{m0}$ and $w_0$, we derived the
marginalized $1\sigma$ constraint $w_a=0.12^{+0.16}_{-0.15}$.
We show the marginalized $1\sigma$ and $2\sigma$ contours of $\Omega_{m0}$ and $w_0$, and $w_0$ and $w_a$ in Fig. \ref{fswlcpl}.
By using the correlation between  $\Omega_{m0}$ and $w_0$, we reconstruct the evolutions of $w(z)$ and $Om(z)$ in Fig. \ref{fswlcpl}.

\begin{figure}[htp]
\centerline{\includegraphics[width=0.6\textwidth]{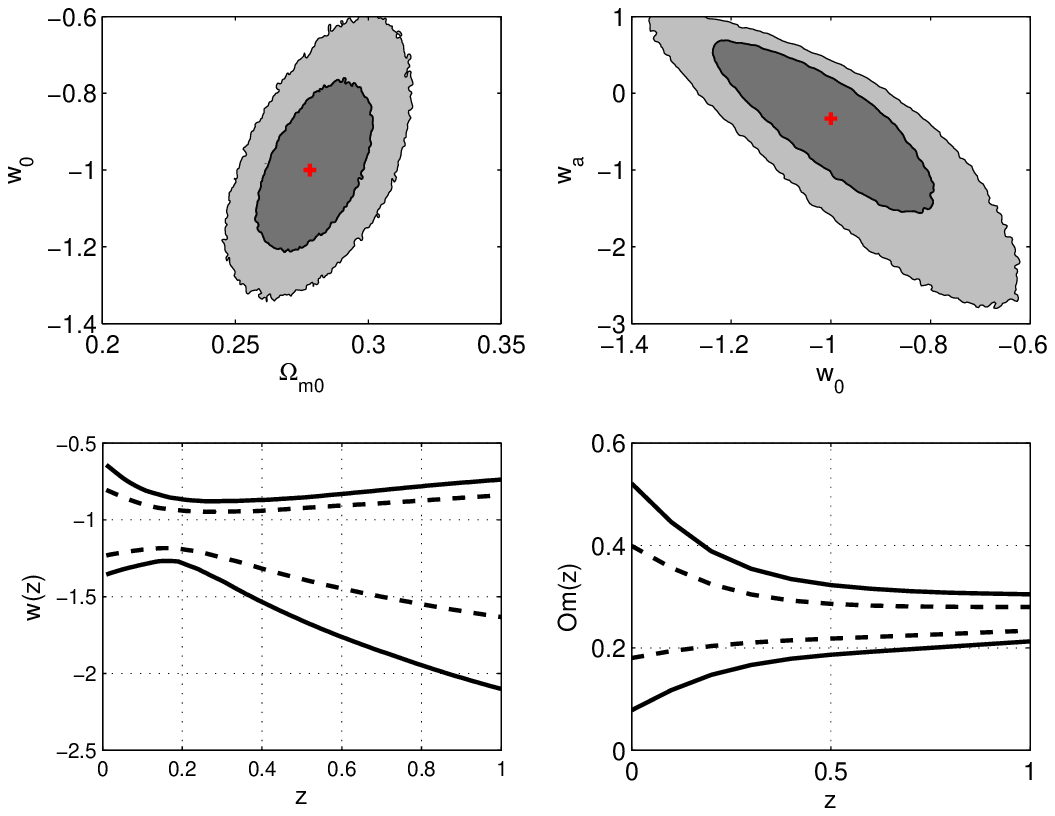}}
\caption{The marginalized $1\sigma$ and $2\sigma$ constraints on the flat CPL model. The upper panels
are for the $1\sigma$ and $2\sigma$ contour plots of $\Omega_{m0}-w_0$ and $w_0-w_a$. The lower panels are the reconstructions
of $w(z)$ and $Om(z)$ by using the constraints on $\Omega_{m0}$, $w_0$ and $w_a$. \label{fcpl}}
\end{figure}

Fitting the CPL model to the observational data, we
get the marginalized $1\sigma$ constraints $\Omega_{m0}=0.278^{+0.018}_{-0.011}$, $w_0=-1.0^{+0.17}_{-0.13}$
and $w_a=-0.33_{-1.03}^{+0.53}$ with $\chi^2=432.4$.
We show the marginalized $1\sigma$ and $2\sigma$ contours of $\Omega_{m0}$ and $w_0$, and $w_0$ and $w_a$ in Fig. \ref{fcpl}.
By using the correlations among the parameters, we reconstruct the evolutions of $w(z)$ and $Om(z)$ in Fig. \ref{fcpl}.
These results are summarized in Table \ref{table1}.

\begin{figure}[htp]
\centerline{\includegraphics[width=0.4\textwidth]{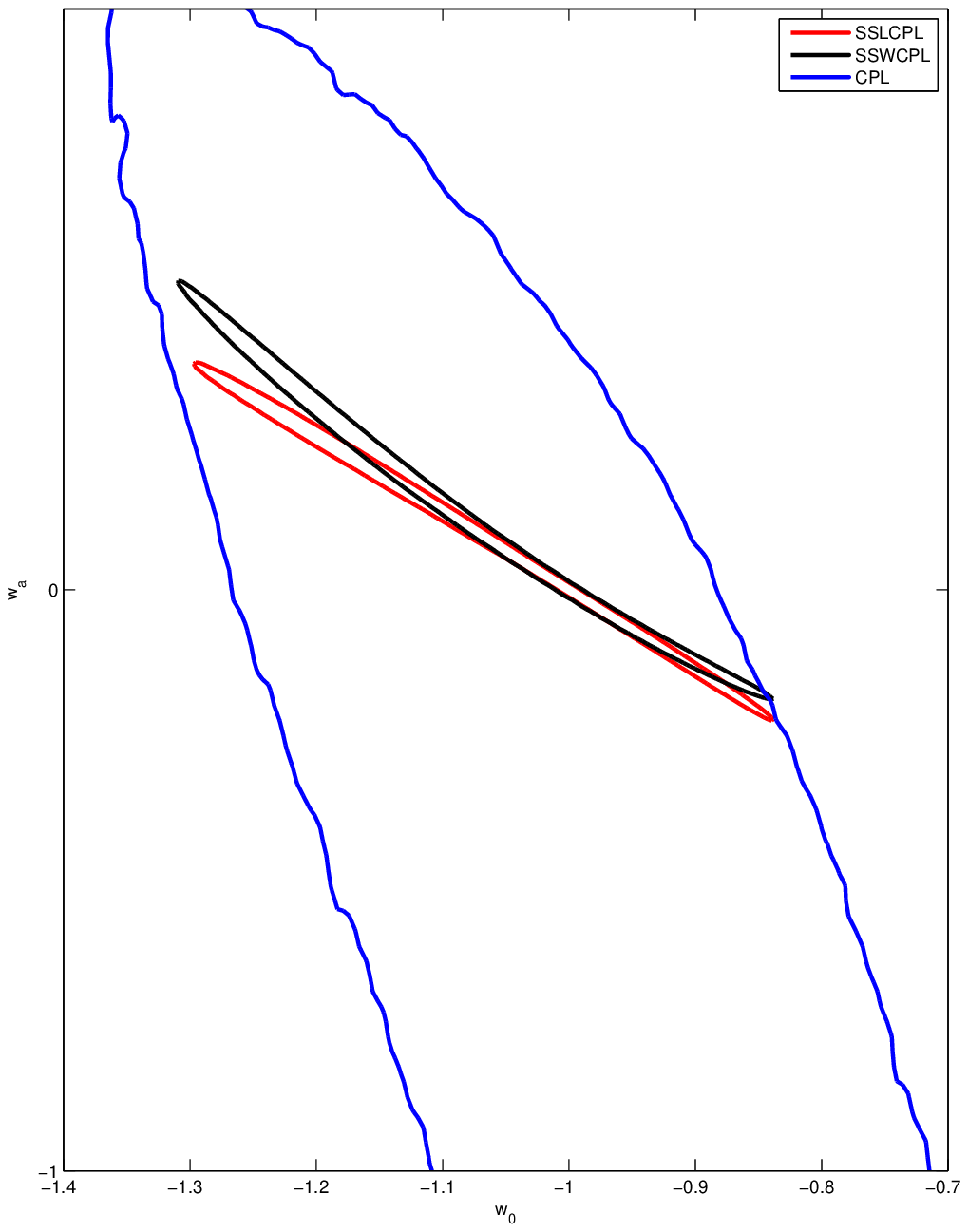}}
\caption{The marginalized $2\sigma$ contours of $w_0$ and $w_a$ for the SSLCPL, SSWCPL  and CPL models.
\label{w0acont}}
\end{figure}

Before comparing the constraints for the SSLCPL and SSWCPL models with the familiar CPL model,
we need to check the consistencies of the degeneracy relations (\ref{waeq1}) and (\ref{twa}).
So we put the $2\sigma$ contour plots of $w_0$ and $w_a$ from Figs. \ref{fsslcpl}-\ref{fcpl}
together in Fig. \ref{w0acont},
we see that the analytical degeneracies  (\ref{waeq1}) for the SSLCPL model and (\ref{twa}) for the SSWCPL model
are consistent with the $w_0-w_a$ contour for the CPL model obtained from the observational constraints,
and the variation of $w(z)$ ($w_a$) is constrained much tighter with the help of analytical relations (\ref{waeq1}) and (\ref{twa}).
The effects of the degeneracy relations (\ref{waeq1}) and (\ref{twa}) on $\Omega_{m0}$ are minimal.
The $1\sigma$ errors of $w_0$ for SSLCPL and SSWCPL models are reduced
around 30\% with the degeneracy relations (\ref{waeq1}) and (\ref{twa})
compared with that in CPL model. Comparing the results from the three models, we find that all three
models fit the observational data well because they give almost the same value of $\chi^2$. In terms of the Akaike
information criterion (AIC) \cite{aic} or Bayesian information criterion (BIC) \cite{bic}, the SSLCPL and SSWCPL model fit the observational
data a little better than the CPL model does. All three models are consistent with $\Lambda$CDM model at the $1\sigma$ level,
as shown explicitly by the $w(z)$ and $Om(z)$ plots in Figs. \ref{fsslcpl}-\ref{fcpl}.

\begin{figure}[htp]
\centerline{\includegraphics[width=0.5\textwidth]{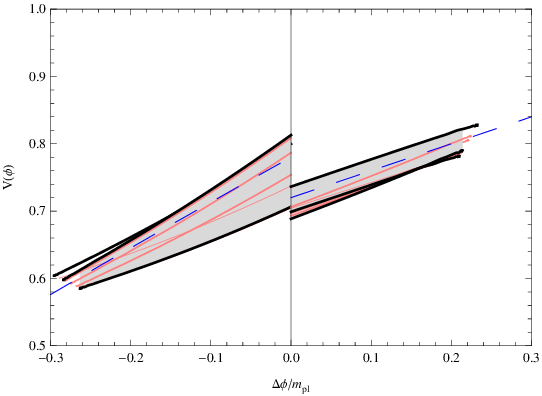}}
\caption{The $1\sigma$ constraint on the thawing potential.
The left side if for the phantom field and the right side is for the quintessence field.
The potential is in the unit of the current critical density $\rho_{cr0}$ and $\Delta \phi$
is the difference between the scalar field $\phi$ and its current value $\phi_0$. The dashed lines
show the potential $V(\phi)=\phi^{0.85}$ for the phantom case and $V(\phi)=\phi$ for the quintessence case.
\label{vphi}}
\end{figure}

With the constraints on the model parameters $\Omega_{m0}$ and $w_0$, we can get the constraints on the forms of the
thawing potential $V(\phi)$ by using the following relations,
\begin{equation}
\label{phieq1}
\left(\frac{d\phi}{d\ln a}\right)^2=3m^2_{pl}\Omega_\phi(a)|1+w(a)|,
\end{equation}
\begin{equation}
\label{vphieq1}
V(a)=\frac{1}{2}\rho_{cr0}(1-w(a))\Omega_\phi(a)(H(a)/H_0)^2,
\end{equation}
where $m_{pl}=(8\pi G)^{-1/2}$ and the current critical density $\rho_{cr0}=3m^2_{pl}H_0^2$.
The allowed $1\sigma$ regions of the thawing potentials for the phantom and quintessence cases are shown in Fig. \ref{vphi}.
For the phantom case, the scalar field climbs up the potential, the potential $V(\phi)=\phi^{0.85}$ shown with the dashed line
is inside the allowed $1\sigma$ region. For the quintessence case, the scalar field rolls down the potential,
the potential $V(\phi)=\phi$ shown with the dashed line is inside the allowed $1\sigma$ region.

\begin{table}[pht]
\caption{The marginalized $1\sigma$ constraints from observational data.}
\begin{tabular}{ccccc}
\hline
Model & $\Omega_{m}$ &  $w_0$ & $w_a$  \\ \hline
flat SSLCPL & $0.275^{+0.015}_{-0.011}$  & $-1.08^{+0.11}_{-0.09}$ & $0.11^{+0.12}_{-0.14}$  \\ \hline
flat SSWCPL & $0.276^{+0.014}_{-0.013}$  & $-1.09\pm 0.10$ & $0.12^{+0.16}_{-0.15}$  \\ \hline
flat CPL & $0.278^{+0.018}_{-0.011}$ &  $-1.0^{+0.17}_{-0.13}$ & $-0.33_{-1.03}^{+0.53}$  \\ \hline
\end{tabular}
\label{table1}
\end{table}

\section{Conclusions}

From the relationship (\ref{eq15}) for thawing models with a nearly flat potential
and the CPL approximation for a wide class of dynamical dark energy models,
we derived the SSLCPL and SSWCPL models which break the degeneracies
among $\Omega_{\phi 0}$, $w_0$ and $w_a$.
The two models reduce to the CPL model with only one free parameter $w_0$.
The relative errors on the equation of state brought by the SSLCPL and SSWCPL approximations are under a few percent,
so these models capture the main dynamics of the thawing model and approximate the dynamics
of a wide class of thawing models well. Instead of studying a particular dynamical dark energy model,
the SSLCPL and SSWCPL models can be used to probe the general properties of dynamical dark energy.
The proposed degeneracy relations for $w_0$ and $w_a$ are consistent with that
found for the familiar CPL model, so the SSLCPL and SSWCPL models are self-consistent.
Current observational data constrain the current value of the equation of state $w_0$ of dark energy around $-1$
with more than 10\% error. The current value of the variation of the equation of state $w_a$ is loosely constrained;
with the relation between $w_0$ and $w_a$ which is proportional to $1+w_0$ found in SSLCPL and SSWCPL models,
$w_a$ is tightly constrained and $w_0\le -1$ at the $1\sigma$ level.
Both models give almost the same minimum $\chi^2$ as the original CPL does when fitting to the observational data.
In terms of AIC or BIC, the models fit the observational data a litter better than the original CPL does.
With the help of relations (\ref{waeq1}) and (\ref{twa}),
the $1\sigma$ error bar of $w_0$ is reduced about 30\%.
The result is almost the same when we replace the WMAP7 data by the Planck data \cite{planck13,wangyun13,Gao:2013pfa}.
Both SSLCPL and SSWCPL models have only one free parameter and they help tighten the constraint on
the property of dark energy.

With the tighter constraints on the parameters $\Omega_{m0}$ and $w_0$, we obtain the constraints
on the thawing potential $V(\phi)$. For the phantom case, the potential $V(\phi)=\phi^{0.85}$ is consistent
with current observations. For the quintessence case, the potential $V(\phi)=\phi$ is consistent with current observations.

\begin{acknowledgements}
This work was partially supported by
the National Basic Science Program (Project 973) of China under
grant No. 2010CB833004, the NNSF of China under grant Nos. 10935013 and 11175270,
the Program for New Century Excellent Talents in University
and the Fundamental Research Funds for the Central Universities.
\end{acknowledgements}


\end{document}